\documentclass[a4paper,twoside]{article}
\pdfoutput=1
\newcommand{\affil}[1]{$^{\rm #1}$}
\usepackage[authoryear]{natbib}
\bibpunct{(}{)}{;}{a}{}{,}
\usepackage{graphicx}
\date{} 
\title{\large\bf\flushleft The Effect of Drag from the Galactic Hot Halo on the Magellanic Stream and Leading Arm}
\author{\parbox{\textwidth}{\flushleft
\vspace{-0.5cm}
{\it Jonathan Diaz\affil{A,B} and Kenji Bekki\affil{A}}\\
\vspace{0.4cm}
{\small \affil{A}\,International Centre for Radio Astronomy Research, M468, The University of Western Australia, 35 Stirling Highway, Crawley, Western Australia 6009, Australia }\\
{\small \affil{B}\,Email: jonathan.diaz@icrar.org}}}
\begin{document}
\twocolumn[
\begin{changemargin}{.8cm}{.5cm}
\begin{minipage}{.9\textwidth}
\vspace{-1cm}
\maketitle
%
%
\small{\bf Abstract:} We study the effect of drag induced by the Galactic hot halo on the two neutral hydrogen (HI) cloud complexes associated with the Large and Small Magellanic Clouds: the Magellanic Stream (MS) and the Leading Arm (LA).  In particular, we adopt the numerical models of previous studies and re-simulate the tidal formation of the MS and LA with the inclusion of a drag term.  We find that the drag has three effects which, although model-dependent, may bring the tidal formation scenario into better agreement with observations: correcting the LA kinematics, reproducing the MS column density gradient, and enhancing the formation of MS bifurcation.  We furthermore propose a two-stage mechanism by which the bifurcation forms.  In general, the inclusion of drag has a variety of both positive and negative effects on the global properties of the MS and LA, including their on-sky positions, kinematics, radial distances, and column densities.  We also provide an argument which suggests that ram pressure stripping and tidal stripping are mutually exclusive candidates for the formation of the MS and LA.

\medskip
\medskip
\end{minipage}
\end{changemargin}
]
\small


\section{Introduction}

The Magellanic Stream (MS) is a massive trail of neutral hydrogen (HI) gas which forms a well-confined arc across more than 100 degrees in the sky.  The Leading Arm (LA), which is composed of a number of discrete HI branches, stretches oppositely to the MS and leads the Large and Small Magellanic Clouds (LMC and SMC, respectively) in their orbit the Milky Way.  There are only two well-studied mechanisms which could have formed the MS and LA by removing gas from the LMC and/or SMC: ram pressure stripping from the Galactic hot halo (e.g., Meurer et al. 1985;  Heller \& Rohlfs 1994; Mastropietro et al. 2005), or tidal stripping from the gravitational interaction between the LMC, SMC, and Milky Way (e.g. Murai \& Fujimoto 1980; Gardiner \& Noguchi 1996, hereafter GN96; Connors et al. 2006).

Both scenarios have their strengths and weaknesses, but because the existence of the LA favors a tidal origin (Putman et al. 1998), the tidal models appear more capable of reproducing the global morphology and kinematics of the system as a whole.  The pure tidal scenario is nevertheless naive because it ignores the effect of the Galactic hot halo \emph{after} the formation of the MS and LA.  Regardless of the disputed role that the Galactic hot halo may have played in the formation of the MS and LA, it is widely accepted that the hot halo has influenced the gas subsequent to its removal from the L/SMC disk.  By ignoring these gas-dynamical effects on the MS and LA, the pure tidal scenario is guilty of a potentially major oversight.

There are a host of observations and simulations which corroborate the hydrodynamical interaction of the hot halo with the MS and LA on small scales (see discussion in section 5), but tracing the impact of the hot halo on \emph{global} scales is less straightforward.  That is, it is difficult to asses what impact, if any, the hot halo has made on (1) the morphology, (2) the kinematics, (3) the radial distance profile, and (4) the column density gradient of the MS and LA.  This ambiguity stems from two sources: not only are the properties of the hot halo (e.g. temperature, density) poorly understood, but the formation and evolution of the MS and LA -- and therefore also their past interactions with the hot halo -- are the subject of debate.

In the present study, we assess the impact of the hot halo on the global properties of the MS and LA by studying two specific tidal models: the GN96 model, which is generally considered to be the traditional tidal scenario; and the Diaz \& Bekki (2011, hereafter DB11) model, which is based upon the increased L/SMC velocities suggested by recent proper motion measurements (Kallivayalil et al. 2006).  The LMC and SMC in the GN96 model have much smaller orbital velocities ($\sim$300 and $\sim$250 km s$^{-1}$ at the present-day, respectively) as compared to the DB11 model ($\sim$360 and $\sim$330 km s$^{-1}$), but the new proper motion estimates of Vieira et al. (2010) imply that both models are valid albeit at opposite extremes of the 1-$\sigma$ limit.  Accordingly, the present study covers two extreme cases for the formation and evolution of the MS and LA: a low velocity model (GN96) and a high velocity model (DB11).

As described in section 2, our method is to re-simulate the GN96 and DB11 models with the insertion of a drag term into the equations of motion.  We construct this drag to be proportional to the ram pressure that would be experienced by the MS and LA as they plunge through the hot halo.  In this way, we can compare the pure tidal model with a ``tidal plus drag" model, and thereby isolate any global effects of the hot halo.  While the GN96 and DB11 models are successful in reproducing many global properties of the MS and LA, they share a common fault (as with every other pure tidal model to date): the predicted line-of-sight velocities of the LA are significantly greater than observed, by as much as $\sim$150 km s$^{-1}$ (Bruns et al. 2005).

In fact, our original motivation in undertaking the present study was to determine whether a drag force induced by the hot halo could bring the predicted LA kinematics into better agreement with observations.  This is a particularly interesting problem in the specific context of the GN96 and DB11 models, because their predicted LA's exhibit morphological and kinematical differences.  As discussed in section 3, we find that drag is indeed able to reduce the discrepancy with observation for both models, although it is accompanied by an assortment of other effects.  For instance, even though the LA kinematics improve, the on-sky position changes.  We discuss these various effects in the context of the ``best" models we could find, both for the GN96 model (section 3.1) and the DB11 model (section 3.2).  Not surprisingly, the best drag model for GN96 requires a different hot halo parameterization than the best model for DB11.

Although we do not present an extensive parameter study for different configurations of the hot halo, we provide a brief discussion in section 4.  In particular, we determine that our parameter values for the ``best" models are approximately an order of magnitude smaller than the values at which the L/SMC disks are altered by drag.  Unfortunately, our parameterization of the hot halo does not permit an exact determination of gas density (see section 2), but we use the ram pressure stripping model of Mastropietro et al. (2005) to suggest what densities might be implied by our parameters.  As a consequence, we conclude that ram pressure stripping cannot occur at the hot halo densities utilized in our best models.  

Also in section 4, we determine the parameter values at which the MS and LA are unable to survive against drag: the LA sinks to the Galactic center, and the MS remains confined in the SMC disk.  Interestingly, these values are comparable but slightly smaller than the values required for altering the disks by drag.  Accordingly, we suggest that gaseous features having a tidal origin cannot survive at the large hot halo densities required for ram pressure stripping.  In other words, we propose that tidal stripping and ram pressure stripping are mutually exclusive candidates for the formation of the MS and LA.

Though the present work focuses on two specific tidal models for the origin of the MS (i.e., GN96 and DB11), it should be noted that there exist other possible models within the tidal formation scenario.  In particular, Besla et al. (2010) argue that the MS may have been formed during a first infall orbit, in contrast to the tightly bound orbits of GN96 and DB11.  Similar to GN96 and DB11, the Besla et al. (2010) model does not include interactions with the Milky Way's hot halo, although SPH gas dynamics is employed for the LMC and SMC disks.  We do not include a first infall scenario into the present study for several reasons: first, hot halo interactions in a first infall orbit will occur over shorter timescales and would therefore have a diminished effect; second, the bifurcation of the MS is not reproduced in the Besla et al. (2010) model, but it is an important property of the DB11 model which is sensitive to the presence of drag, as described in section 3.2; and third, the on-sky position of the LA in the Besla et al. (2010) model is incorrect, which would undermine an analysis of the effect of drag on LA kinematics.



\section{Numerical Method}

For detailed descriptions of the numerical models that we presently consider (GN96 and DB11), we refer the reader to the original papers.  Each of the models is composed of two components: first, backwards orbit integration with constraints given by present-day orbital parameters; and second, N-body evolution of the SMC disk on the previously computed orbit.  It should be noted that the original DB11 model is based on \emph{test particles}, but in this study we have extended the model to an N-body approach.  Interestingly, the test particle version (original DB11 paper) and the N-body version (this study) are only marginally different in their global properties.  Our re-simulation of the GN96 model is identical to the original version, except for having a higher mass resolution.  The re-simulations of GN96 and DB11 will hereafter be referred to as the \emph{pure tidal models}.

Our pure tidal models are based on an N-body approach similar to those adopted in previous studies of the evolution of the SMC (e.g., Connors et al. 2006; Bekki \& Chiba 2009).  Since the details of our numerical method are given in our previous papers (e.g., Bekki \& Chiba 2009), we describe it only briefly here.  We consider that the SMC has a collisionless disk and a dark matter halo before tidal interaction with the LMC and Milky Way.  We construct the disk to mimic the distribution of gas, and even though the disk particles are collisionless, we hereafter refer to them as ``gas particles."  The disk follows an exponential profile with a scale-length of 1.5 kpc and a truncation radius of 7.5 kpc, and the total number of particles is $10^5$ with a total mass of $1.5 \times 10^9 {\rm M}_{\odot}$.

The dark matter halo is also composed of $10^5$ collisionless particles having a total mass of $1.5 \times 10^9 {\rm M}_{\odot}$, and it follows an NFW profile truncated at 7.5 kpc (Navarro et al. 1996).  The gravitational softening length in each simulation is set to be 110 pc.  Gas dynamics, star formation processes, and chemical evolution included in our previous simulations of the LMC and SMC (Bekki \& Chiba 2005; 2009) are not included in the present study.  In our future papers, we will incorporate these three physical processes into a more sophisticated simulation in order to discuss how they each play a role in the formation of the MS.  However, our present interest is limited to understanding how drag from the Galactic hot halo can change the results of the pure tidal models.

Our next step is to change the dynamics of the pure tidal models by incorporating a drag term.  The deceleration due to drag is proportional to the ram pressure that the hot halo exerts on the orbiting gas clouds (e.g., MS and LA).  This ram pressure is expressed as
\begin{equation} \label{rp}
P = \rho v^2,
\end{equation}
where $\rho$ is the density of the hot halo, and $v$ is the magnitude of the relative velocity between the gas cloud and the hot halo (Gunn \& Gott 1972).

The density profile of the hot halo is poorly understood, which in principle would allow considerable freedom in defining $\rho$, but imposing hydrostatic equilibrium between the hot halo and the underlying dark matter would reduce the range of possible profiles for $\rho$.  In the fully hydrodynamical simulations of Crain et al. (2010), the gradient of hot halo density is observed to be steeper than that of the underlying dark matter profile, which in turn suggests that hydrostatic equilibrium may not necessarily imply a strict match of density profiles.  For convenience we choose to match our hot halo density to that of the Galactic dark matter distribution of the pure tidal models.  Gardiner (1999) has previously studied the effect of drag on the GN96 model, but his adopted hot halo follows a markedly different profile than the dark matter halo, which is improbable though not impossible.

In GN96 and DB11, the Milky Way is assigned a logarithmic potential of the form $\phi=-V_{\rm cir}^2\ln(r)$, with $V_{\rm cir}=$ 220 km s$^{-1}$ for GN96 and $V_{\rm cir}=$ 250 km s$^{-1}$ for DB11.  This choice of potential corresponds to an isothermal dark matter halo, for which the density profile is spherically symmetric and falls off as $\sim$r$^{-2}$.  We therefore take the density of the hot halo to be 
\begin{equation}\label{rho}
\rho(r) = \frac{ \rho_o }{ 1 + (\frac{r}{r_c})^2 },
\end{equation}
where $\rho_o$ is the central density and $r_c$ is the core radius (e.g., Westmeier et al. 2010).  Because the hot halo is tenuous, the value of $\rho_o$ is negligibly small when compared to the density of the Galactic dark matter halo.  For this reason, the gravitational influence (i.e., total mass) of the hot halo can be safely ignored.  We may also disregard the core radius $r_c$, because varying $r_c$ within reasonable values has only a marginal effect at the distances of the LMC and SMC ($\sim50-60$ kpc).  For our study, we take $r_c=1.0$ kpc for convenience.

In order to incorporate (\ref{rp}) into the equations of motion, we must know the area $A$ of the cloud which is experiencing the ram pressure.  The deceleration of this cloud will then be
\begin{equation}\label{dec}
\vec{a}(r,\vec{v}) = - \frac{ \alpha }{ 1 + r^2 } v \vec{v},
\end{equation}
where $\alpha$ is a catch-all parameter (Meurer et al. 1985; Heller \& Rohlfs 1994) equal to
\begin{equation}\label{alpha}
\alpha = \frac{  \rho_o A C_D }{ m }.
\end{equation}
The drag efficiency $C_D$ is of order unity, and $m$ designates the mass of the cloud.

To include the effect of drag in our simulations, we may simply insert (\ref{dec}) into the equations of motion for every gas particle.  The magnitude of this drag term depends on the value of $\alpha$, which may be tuned as a free parameter.  Unfortunately, because the simulation utilizes point masses, the cross-sectional area $A$ becomes meaningless.  It is furthermore unreasonable to assume a certain value for $A$, because in reality, the value of $A$ will vary from cloud to cloud depending on local hydrodynamical properties.  Consequently, equation (\ref{alpha}) implies that $\alpha$ is a degenerate quantity: for a given value of $\alpha$, the hot halo density parameter $\rho_o$ cannot be uniquely determined.  Nevertheless, in section 4 we provide an argument on how to understand the value of $\alpha$ in terms of hot halo density.

Because the foregoing methodology is only an approximate scheme for handling the interaction between the MS and the hot halo, a number of important hydrodynamical features are overlooked, such as the presence of bow shocks, cloud ablation, ionization, and phase mixing (Bland-Hawthorn et al. 2007; Heitsch \& Putman 2009).  We suggest that these features affect the MS on a local scale (i.e., affecting individual clouds) in contrast to the global scales which occupy our interest in the present work.  In particular, our methodology adequately captures the evolution of global properties such as the variation of radial distances, column densities, and velocities along the MS and LA.

It is difficult to constrain three dimensional orbits within a self-consistent hydrodynamical model, and although future work may lead to such a model, we presently take the more efficient route of introducing a drag term to approximate the hydrodynamics.  This same method has been adopted previously to treat interactions between the MS and the hot halo (e.g., Meurer et al. 1985; Heller \& Rohlfs 1994; Gardiner 1999).


\section{The Effect of Hot Halo Drag}

\subsection {GN96 Model}

The top panel of Figure 1 shows the orbital separations of the LMC, SMC, and Milky Way in the GN96 model.  The MS and LA are pulled out of the SMC disk $\sim$1.5 Gyr ago, coinciding with a pericentric approach about the Milky Way.  The inclusion of drag does not affect this formation scenario; rather, it affects the subsequent evolution of the MS and LA from $\sim$ 1.5 Gyr ago to the present day.  In Figures 2, 3, 4, and 5, we compare the results of the pure tidal model with the case of incorporating drag.  For our drag model, we have chosen $\alpha = 0.3$.

\begin{figure}
\begin{center}
\includegraphics[width=8cm]{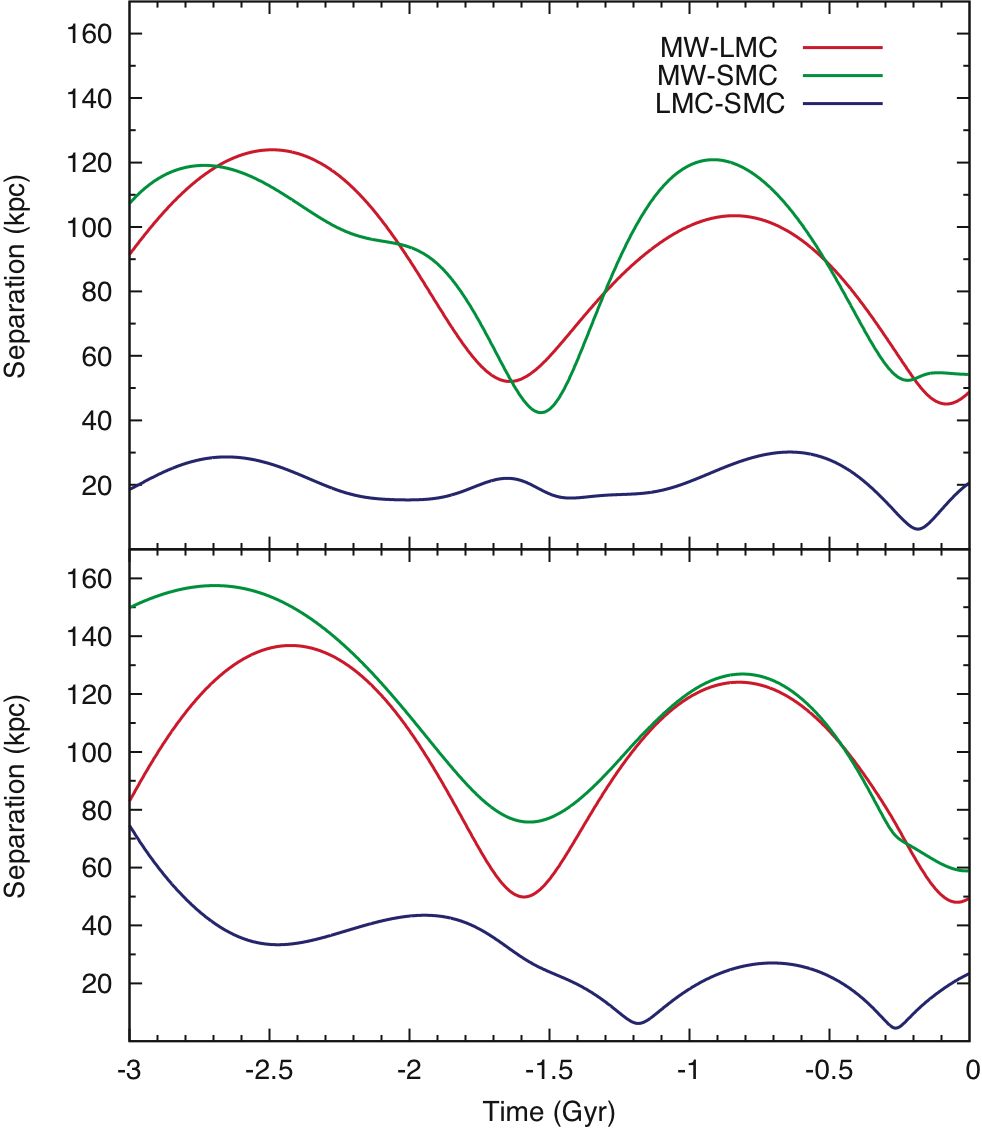}
\caption{Separations between the LMC, SMC, and Milky Way during their 3 Gyr orbital interaction for the GN96 model (top panel) and DB11 model (bottom panel).}\label{}
\end{center}
\end{figure}

From Figure 2, it is clear that the position of the MS does not change in response to drag.  However, the position of the LA changes significantly: the particles at the tip of the LA are dispersed into a long arc which wraps around the sky.  The tip of this drag-induced arc appears to approach the position of the MS around galactic longitude $l=45^{\circ}$, but this is an artificial consequence of the on-sky projection.  We will discuss this point later, but for now we mention briefly that despite the on-sky appearance, the new tip of the LA in the drag model is actually separated from the MS by large distances (see bottom panel of Figure 4).  The observed LA does not have such a dispersed morphology (Bruns et al. 2005), and therefore we consider this drag-induced effect to be undesired.

\begin{figure}
\begin{center}
\includegraphics[width=8cm]{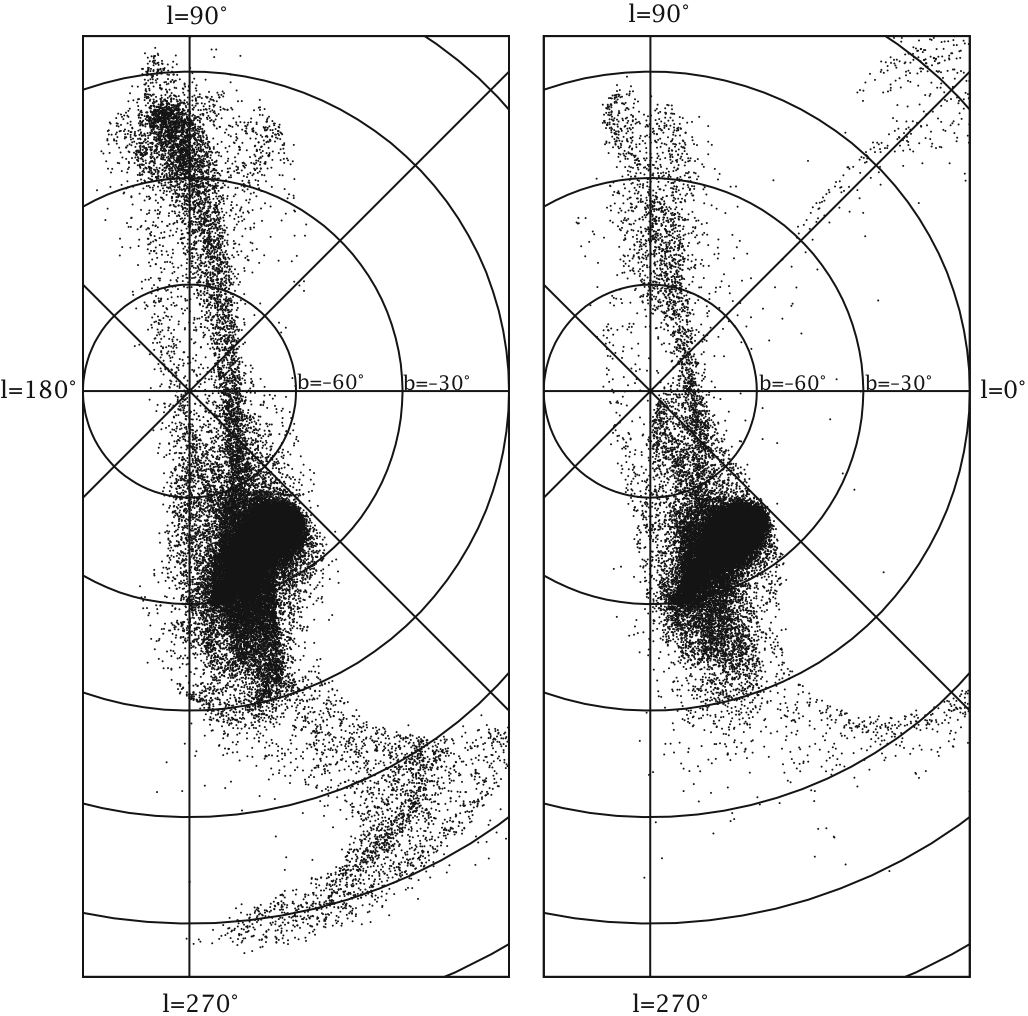}
\caption{Present-day distribution of gas particles for the pure tidal GN96 model (left panel) and the GN96 model with drag (right panel).  The on-sky projection is given in galactic coordinates ($l$,$b$).}\label{}
\end{center}
\end{figure}

In the top panel of Figure 3, we plot the kinematics of the MS and LA in the pure tidal model of GN96, and we compare this against observational data from Bruns et al. (2005).  The MS (at Magellanic longitudes $l_{\rm M} > 0^{\circ}$) follows a descending trend in velocity, whereas the velocities of the LA ($l_{\rm M}< 0^{\circ}$) follow a slightly descending yet comparatively flat distribution.  As seen in the bottom panel of Figure 3, the inclusion of drag does not alter the kinematics of the MS.  In contrast, the kinematics of the LA between $-75^{\circ} < l_{\rm M} < 0^{\circ}$ are significantly improved under the influence of drag.  Whereas the pure tidal model over-predicts the velocities by as much as $\sim 150$ km s$^{-1}$, the drag model provides a very encouraging agreement with observations.  However, beyond this range (i.e., for $l_{\rm M} < -75^{\circ}$), there is an obvious disagreement with observations, as the LA should not extend into this region.

\begin{figure}
\begin{center}
\includegraphics[width=8cm]{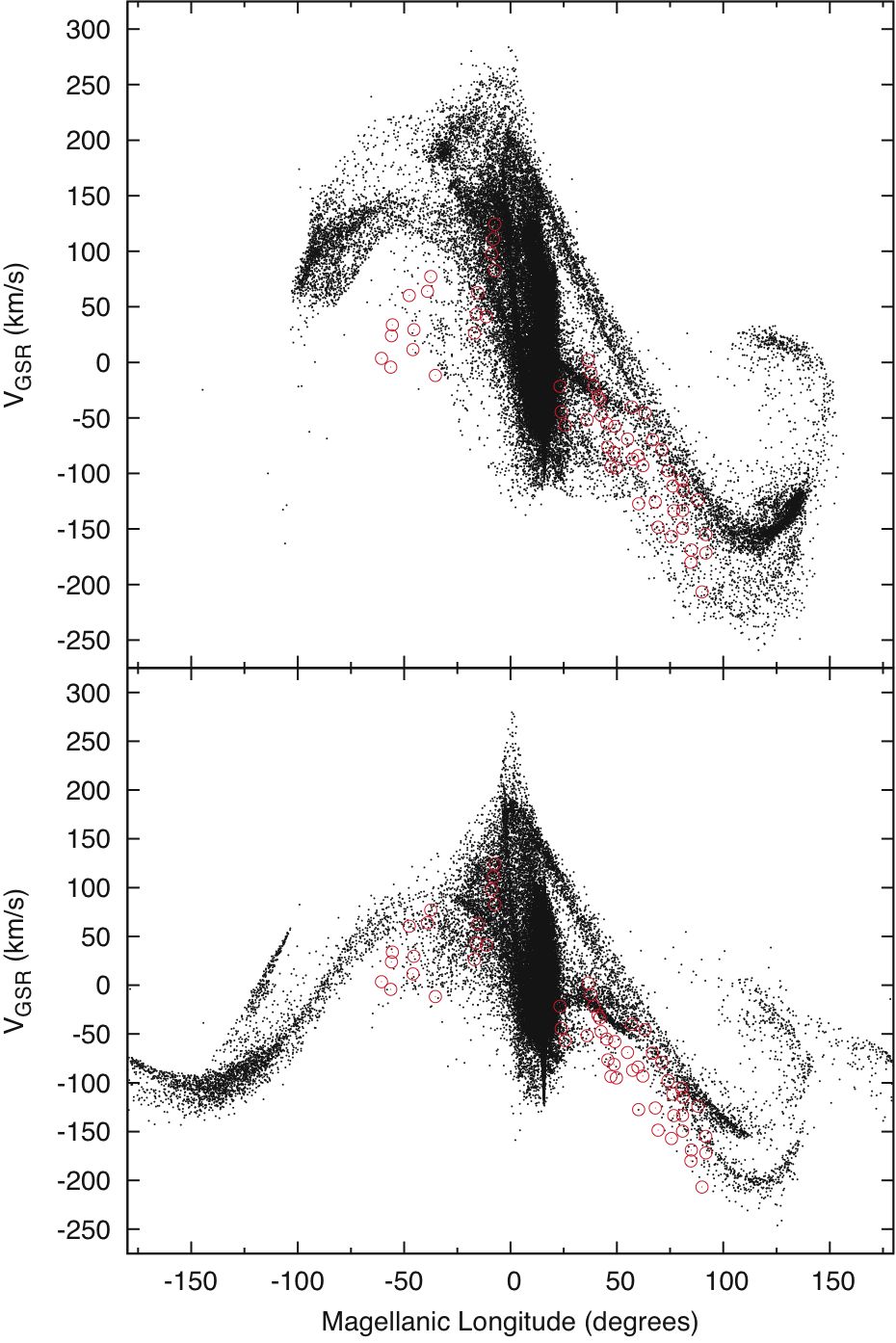}
\caption{Galactocentric radial velocities of gas particles for the pure tidal GN96 model (top panel) and the GN96 model with drag (bottom panel).  The LA is found at Magellanic longitudes $l_{\rm M} < 0$, and the MS at $l_{\rm M} > 0$.  See Wakker (2001) for definition of Magellanic longitude $l_{\rm M}$.  Circles represent observational data sampled from Bruns et al. (2005).}\label{}
\end{center}
\end{figure}

Figure 4 indicates the Galactic radial distances of the particles of the MS and LA as a function of Magellanic longitude.  In the pure tidal model (top panel), the radial distances along the MS increase from $\sim$50 kpc near the SMC ( $l_{\rm M} \approx 0^{\circ}$) to $\sim$200 kpc at the MS tip ( $l_{\rm M} \approx 150^{\circ}$).  The radial distances along the LA also exhibit an increasing trend, from $\sim$40 kpc at the base of the LA to $\sim$75 kpc at its tip ($l_{\rm M} \approx -100^{\circ}$).  The drag model (Figure 4, bottom panel) maintains this increasing trend for the MS, but the change in the LA is dramatic, as the LA is forced to sink to much smaller Galactic radii.  The drag-induced sinking of the LA clarifies that the apparent on-sky proximity to the MS (Figure 2, right-hand panel) is merely a projection effect.  The MS lies at far greater distances ($>40$ kpc) than the LA ($\sim$15 kpc) between $50^{\circ} < l_{\rm M} < 150^{\circ}$.  Due to their large separation, we can assert that the LA and MS arrived at similar locations in the sky ($l \approx 45^{\circ}$) via independent paths of evolution.

Unfortunately, there are no direct observational methods available which can determine distances along the MS and LA.  Consequently, we cannot fully assess whether the inclusion of drag produces a ``positive" effect on the predicted radial distances of the pure tidal models.  However, one data point (which is not shown in Figure 4) does exist: McClure-Griffiths et al. (2008) determined the kinematic distance to one cloud of the LA which has impacted the Galactic HI disk.  They determined a Galactic distance of $\sim 17$ kpc for the cloud, which should be located around $l_{\rm M} \approx -16^{\circ}$.  This distance is significantly less than the predicted distances of the pure tidal model, but because the inclusion of drag makes the LA sink to lower radii, there is a promising indication that drag may reduce the discrepancy with observation.

\begin{figure}
\begin{center}
\includegraphics[width=8cm]{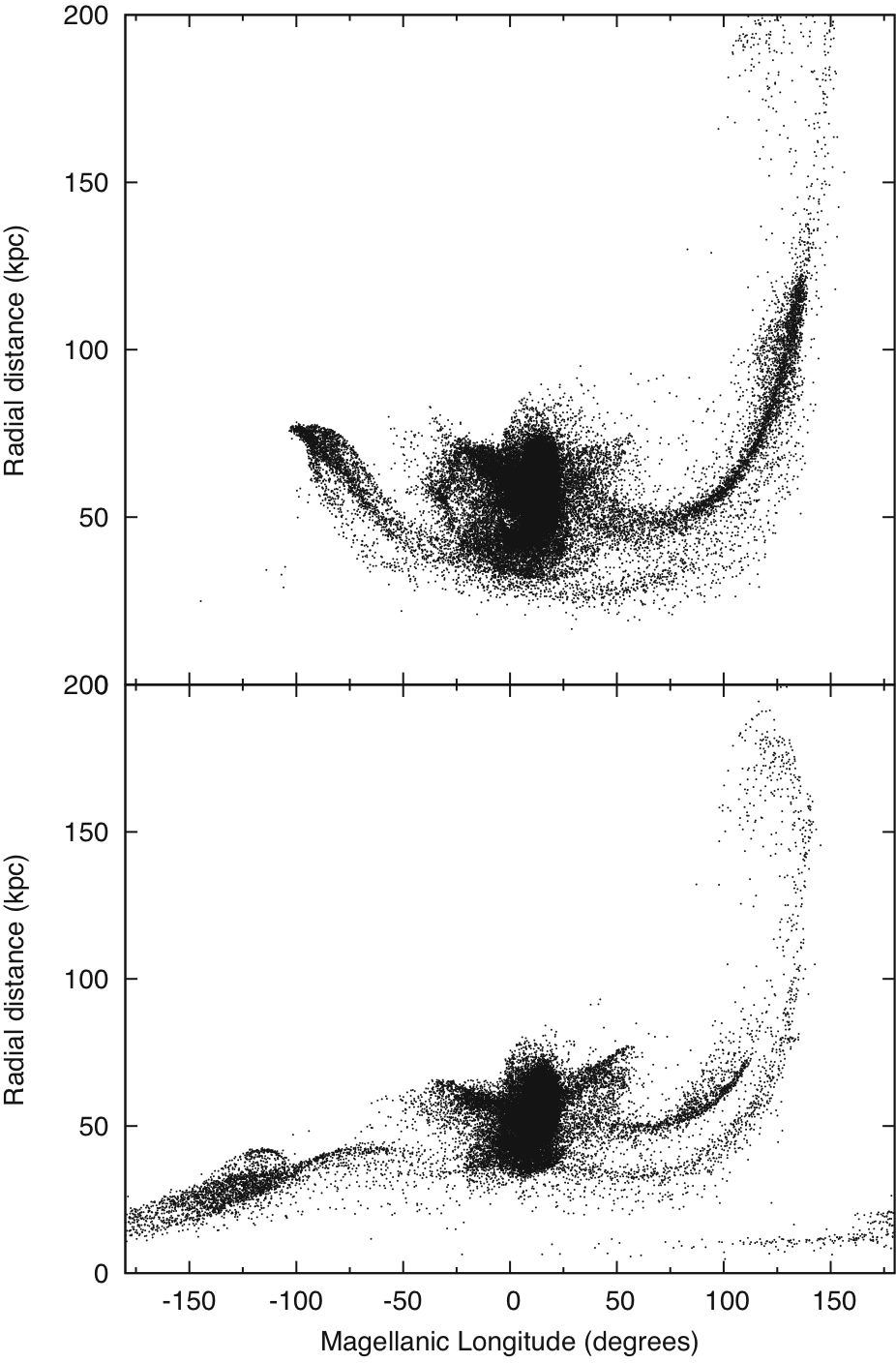}
\caption{Radial distances to the gas particles of the pure tidal GN96 model (top panel) and the GN96 model with drag (bottom panel).  The distances are calculated in a galactocentric frame.  The LA is found at Magellanic longitudes $l_{\rm M} < 0$, and the MS at $l_{\rm M} > 0$.}\label{}
\end{center}
\end{figure}

Even though drag does not alter the trend in radial distance for the MS, there is nevertheless a tangible effect on the MS in Figure 4.  In the drag model (bottom panel), the MS contains three distinct radially separated filaments.  Two of the filaments are obvious, and the third is located at a distance of $\sim$70 kpc between $25^{\circ} < l_{\rm M} < 50^{\circ}$.  The filaments also exist in the pure tidal case (top panel), but the presence of drag allows these structures to become more readily identifiable.  We stress that the hot halo drag does \emph{not} create these MS filaments.  Instead, the drag plays a supplemental role by reducing the velocity dispersion of the MS, which then allows self-gravity to enhance the growth and separation of the filaments.  We find that the MS velocity dispersion changes by $\sim$14.0\% under the influence of drag, which is small enough to retain the overall morphology of the MS and yet large enough to encourage the growth of distinct filaments.

In Figure 5 we plot the number of particles in the MS and LA as a function of Magellanic longitude.  The pure tidal model (solid line) exhibits two pronounced peaks in the number of particles, one at the tip of the MS and another at the tip of the LA.  The peak at the tip of the MS ($100^{\circ} < l_{\rm M} < 150^{\circ}$) strongly disagrees with observations.  The column density of the MS (which is the observational analogue of ``particle count") is actually observed to be gently declining from its base to its tip (Putman et al. 2003a), meaning that the tip of the MS should have a low particle count.  Indeed, the prediction of a plume of particles at the MS tip is one of the main criticisms against the GN96 model.

\begin{figure}
\begin{center}
\includegraphics[width=8cm]{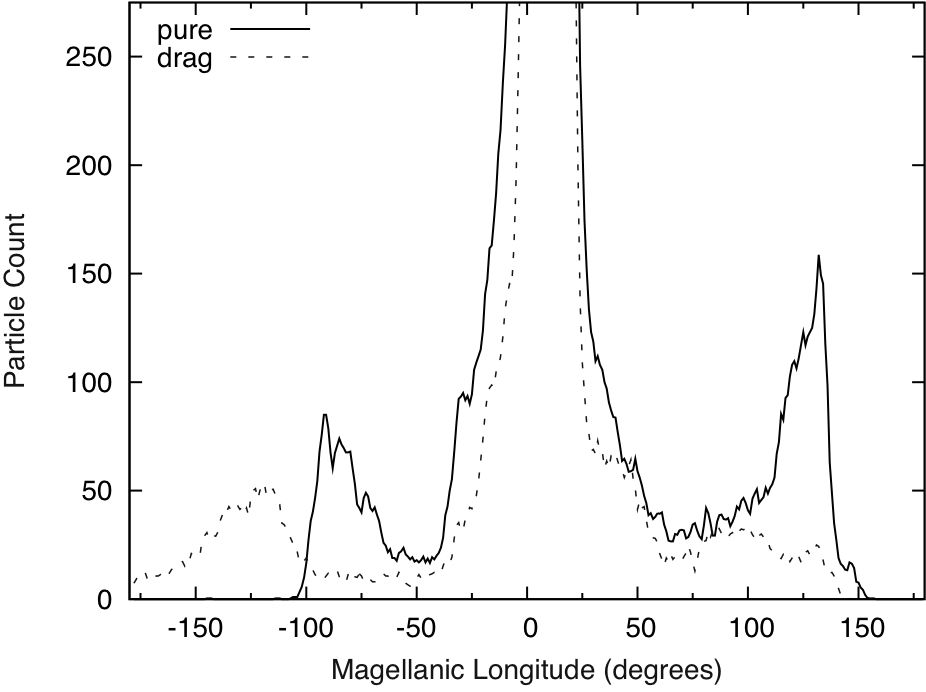}
\caption{Total number of particles in the pure tidal GN96 model (solid line) and the GN96 model with drag (dashed line) plotted as a function of Magellanic longitude.}\label{}
\end{center}
\end{figure}

The MS particle count changes significantly under the influence of drag (Figure 5, dashed line): the peak at the MS tip is completely removed, and the profile along the MS becomes gently declining.  The inclusion of drag therefore resolves an unwanted feature of the GN96 model and improves the agreement with observation.  The drag is able to achieve this by shortening the main filament of the MS (i.e., the highest density filament in Figure 4).  Comparing the top and bottom panels of Figure 4 clearly reveals this shortening.  Drag is also able to reduce the peak at the tip of the LA, though only slightly.  It is shifted to a different position on the sky, which, as stated earlier, does not correspond to any observed HI features of the Magellanic system.

\subsection{DB11 Model}

The bottom panel of Figure 1 shows the orbital separations of the SMC, LMC, and Milky Way in the DB11 model.  The orbits of the LMC and SMC appear similar to those of GN96 (top panel of Figure 1), especially because a pericentric passage about the Milky Way $\sim$1.5 Gyr ago occurs in both models.  But in the case of DB11, this close encounter does \emph{not} contribute to the removal of the MS and LA from the SMC disk.  Instead, the recent formation of a strong LMC-SMC binary pair $\sim$1.2 Gyr ago provides the responsible stripping mechanism.  Additionally, the SMC and LMC have greater orbital energies in the DB11 model due to the adoption of recent proper motions (Kallivayalil et al. 2006; Vieira et al. 2010).  A bound orbit is retained by virtue of assigning a greater circular velocity and therefore greater mass to the Milky Way.

In Figures 6, 7, 8, 9, and 10 we compare the properties of the DB11 pure tidal model with our best ``tidal plus drag" model.  For the inclusion of drag we choose $\alpha =$0.6, which is twice the value chosen for the case of GN96.  It is not surprising that the DB11 model requires a greater value for $\alpha$, because the initial kinetic energies of the MS and LA are greater in DB11 as compared to GN96.  This fact is attributed to the difference in formation mechanisms and also the difference in orbital energies of the SMC.  Consequently, a greater drag force must be employed in order to significantly alter the positions, kinematics, etc. of the MS and LA, which explains the need for a larger value of $\alpha$.

Figure 6 gives the on-sky projection of the MS and LA in the pure tidal model (left-hand panel) and in the model which incorporates drag (right-hand panel).  Figure 7 gives a zoomed-in view of the MS in each of these models.  From these figures, we can discern a variety of both positive and negative effects of drag.  An obvious negative effect is the shortening of the MS, which is particularly worrisome because the MS in the DB11 pure tidal model is already too short ($\sim$100$^{\circ}$) in comparison with the most recent observational estimate ($\sim 140 ^{\circ}$; Nidever et al. 2010).  The inclusion of drag only worsens the situation by shortening the MS by a further $\sim20^{\circ}$.  Another obvious effect of the drag is the destruction of two tenuous structures.  One of the LA branches is unable to survive against drag, which of course is undesired.  However, the drag also destroys a thin strand of particles running parallel to the MS, which must be considered an improvement of the model because the strand does not correspond to any observed HI features.

\begin{figure}
\begin{center}
\includegraphics[width=8cm]{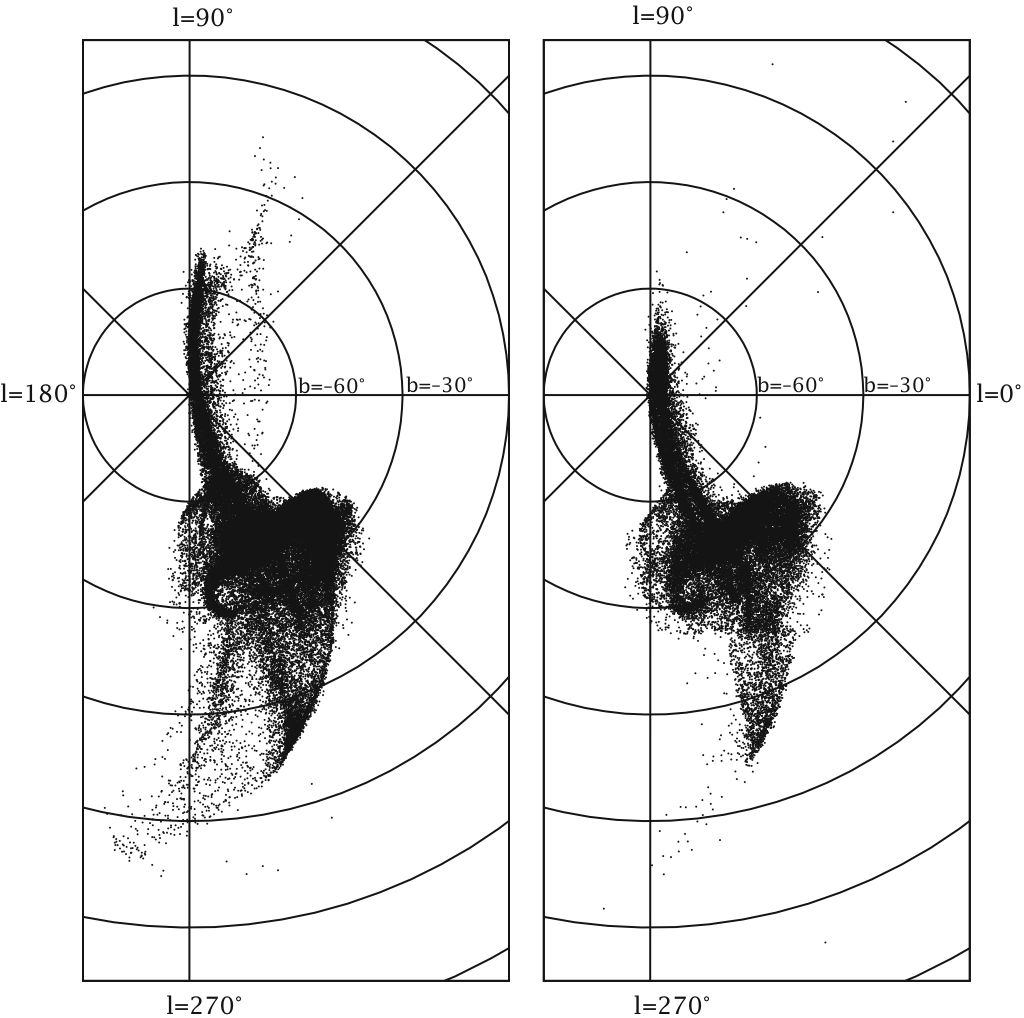}
\caption{Present-day distribution of gas particles for the pure tidal DB11 model (left panel) and the DB11 model with drag (right panel).  The on-sky projection is given in galactic coordinates ($l$,$b$).}\label{}
\end{center}
\end{figure}

\begin{figure}
\begin{center}
\includegraphics[width=8cm]{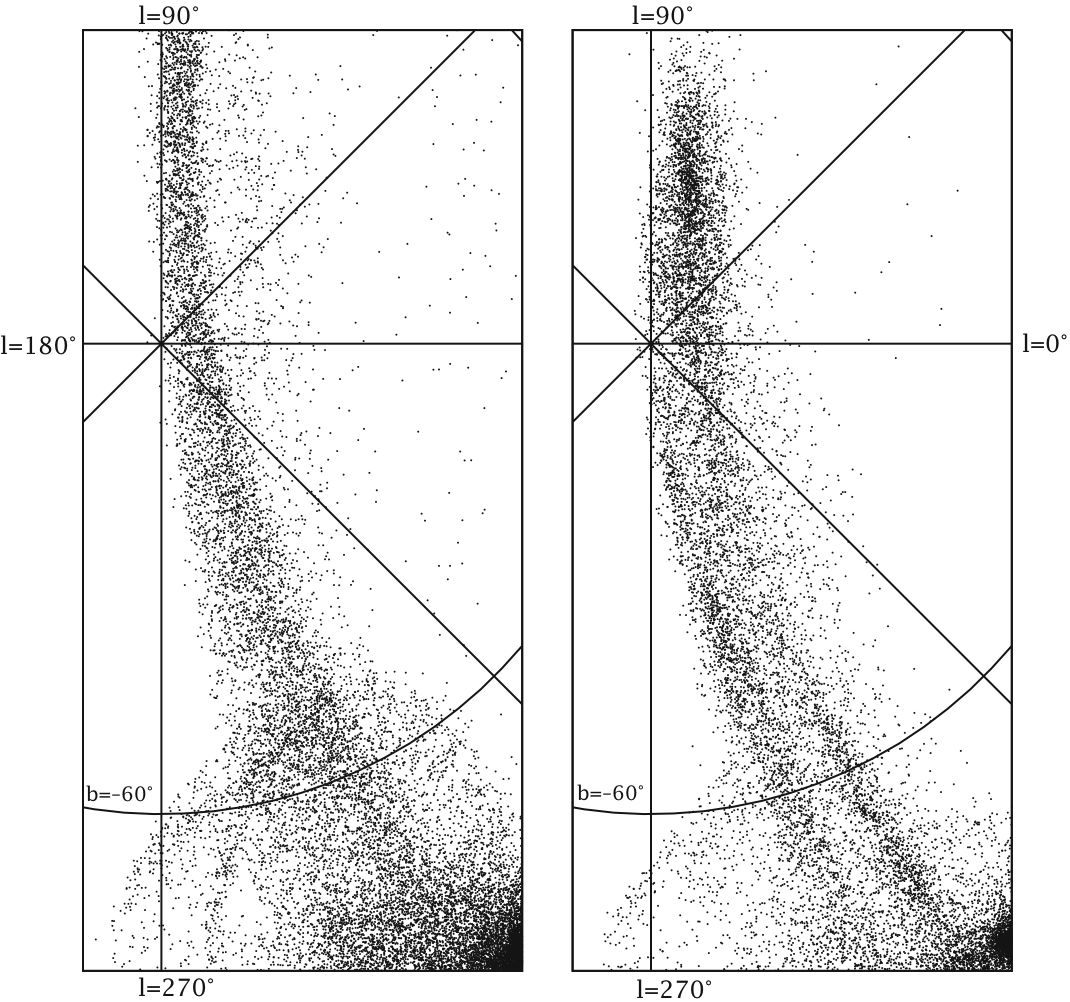}
\caption{The same as Figure 6 but showing only the MS.  The MS exhibits a clear bifurcation in the DB11 model with drag (right panel).}\label{}
\end{center}
\end{figure}

Perhaps the most compelling aspect of the DB11 drag model is the prominent bifurcation of the MS (highlighted in Figure 7, right panel), which corresponds very closely to the position and extent of the observed MS bifurcation (Putman et al. 2003a).  We stress that the hot halo drag is  \emph{not} responsible for creating the bifurcated filaments.  Instead, we propose the following two-stage formation scenario for the MS bifurcation: first, filamentary structure within the MS is created through tidal interactions alone (see original DB11 paper); and second, drag is able to reduce the velocity dispersion within the MS (by $\sim17.4 \%$ for DB11), which enhances the growth and separation of the filaments under self-gravity.   Whether the enhanced filamentary structure of the MS is able to appear as an on-sky bifurcation becomes a model-specific issue.  In the GN96 model for instance, drag is able to enhance the MS filaments without producing a clear bifurcation.  On the other hand, essentially the same physical mechanisms are able to convincingly reproduce the MS bifurcation within the DB11 model.

Figure 8 shows the kinematics of the MS and LA as a function of Magellanic longitude $l_{\rm M}$, and a comparison with observational data is provided by the red circles (Bruns et al. 2005).  The kinematics of the MS ($l_{\rm M} > 0^{\circ}$) in the pure tidal model (top panel) exhibit a strong agreement with the data, except for a single tenuous strand which sticks out horizontally from the main body of the MS.  As stated previously, this strand does not correspond to any observed HI features of the system.  As desired, the hot halo drag is able to destroy this strand of particles, which can be verified in the bottom panel of Figure 8.  Other than this, the kinematics of the MS are largely unaltered by drag, despite the aforementioned shortening of the MS.

\begin{figure}
\begin{center}
\includegraphics[width=8cm]{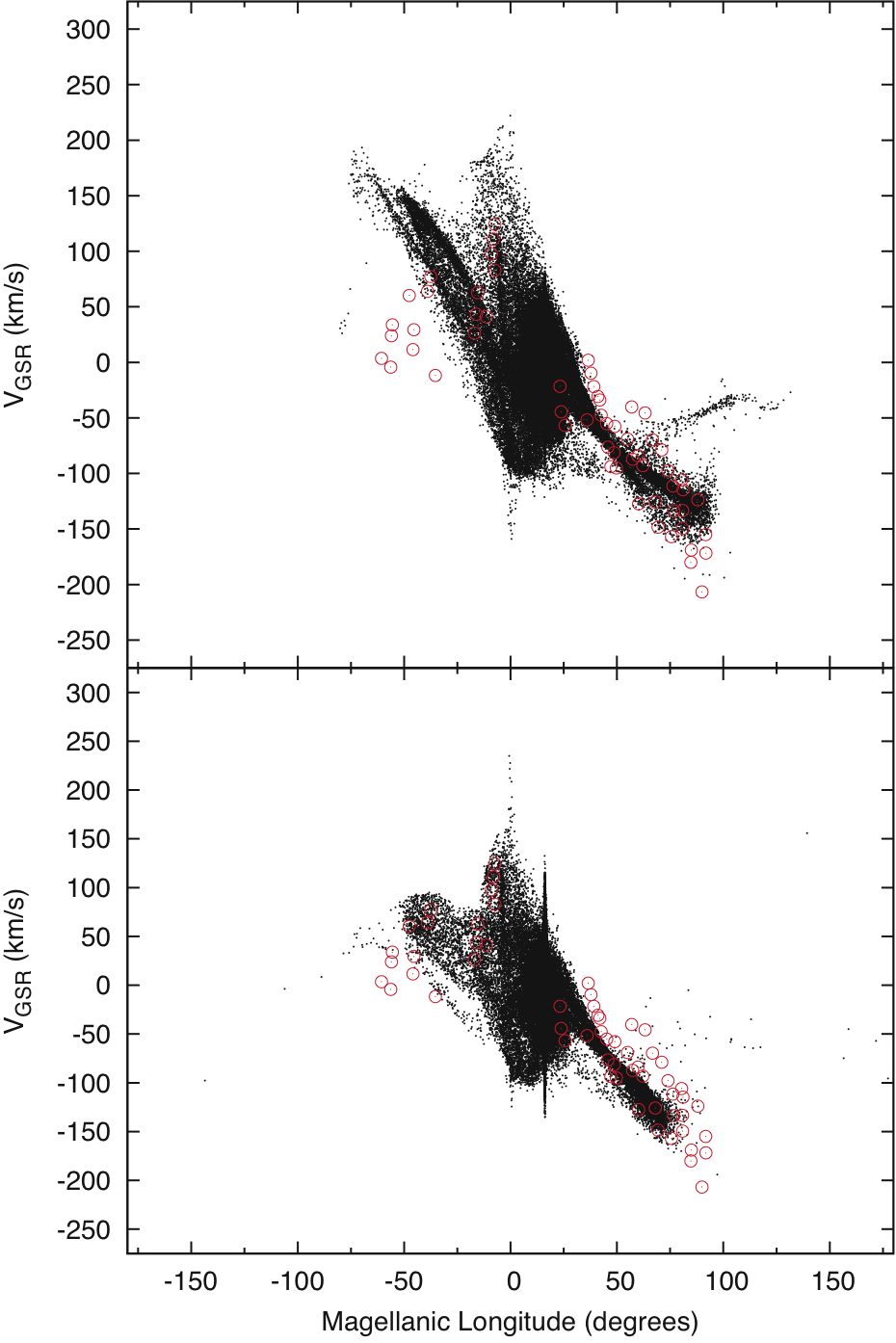}
\caption{Galactocentric radial velocities of gas particles for the pure tidal DB11 model (top panel) and the DB11 model with drag (bottom panel).  The LA is found at Magellanic longitudes $l_{\rm M} < 0$, and the MS at $l_{\rm M} > 0$.  See Wakker (2001) for definition of Magellanic longitude $l_{\rm M}$.  Circles represent observational data sampled from Bruns et al. (2005).}\label{}
\end{center}
\end{figure}

The kinematics of the LA ($l_{\rm M} < 0^{\circ}$) improve significantly in response to drag.  The LA of the pure tidal model (top panel of Figure 8) exhibits a sharply rising trend in contrast to the observed profile which is flat or perhaps slightly decreasing.  The disagreement with observation is severe, with a discrepancy of as much as $\sim$150 km s$^{-1}$.  The drag is able to decrease these LA velocities and recover a strong agreement with observations between $-50^{\circ} < l_{\rm M} < 0^{\circ}$ (Figure 8, bottom panel).  However, there are a few data points surrounding $l_{\rm M} = -50^{\circ}$ which are not reproduced in the drag model, indicating the need for further improvement of the model.  Additionally, the upward trend in the LA kinematics is made only slightly more shallow in response to drag, whereas a flat profile may be preferred.  Nevertheless, the inclusion of drag provides a considerable improvement on the LA kinematics.

The Galactic radial distances of the MS and LA are plotted as a function of Magellanic longitude in Figure 9.  Unlike the GN96 model, the LA does not sink to considerably lower Galactic radii in response to drag.  The drag does indeed pull the LA in the DB11 model to smaller radii, but the change is on the order of only $\sim$5 kpc.  Much higher values for $\alpha$ are needed to cause the radial distance of the LA to significantly decrease, which puts the model at odds with the observation of McClure-Griffiths et al. (2008; see section 3.1).  The radial profile of the MS is also largely unchanged by drag.  Once again, the most obvious effect discernible in Figure 9 is the shortening of the MS and the destruction of the strand which rises beyond 200 kpc.

\begin{figure}
\begin{center}
\includegraphics[width=8cm]{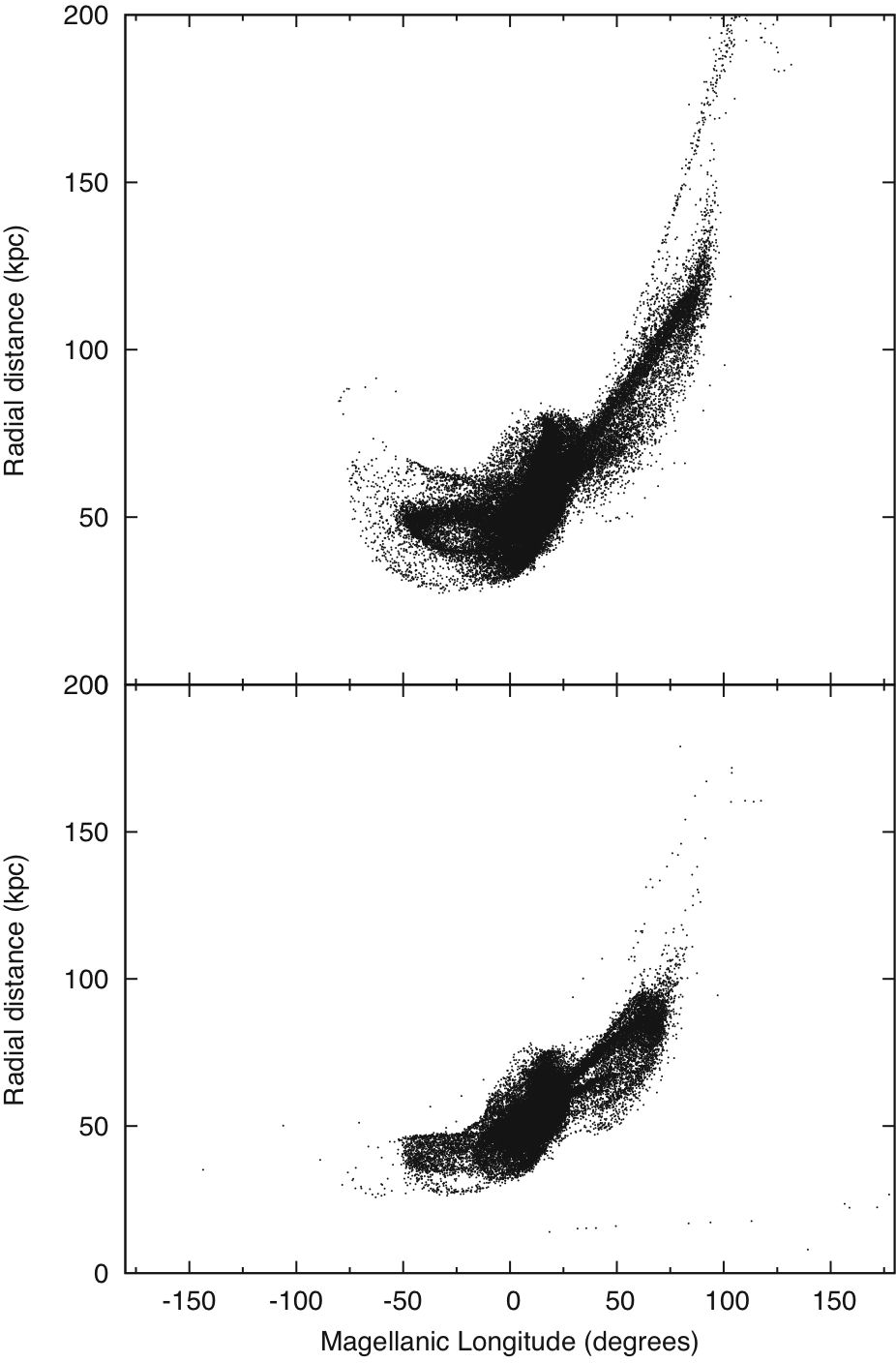}
\caption{Radial distances to the gas particles of the pure tidal DB11 model (top panel) and the DB11 model with drag (bottom panel).  The distances are calculated in a galactocentric frame.  The LA is found at Magellanic longitudes $l_{\rm M} < 0$, and the MS at $l_{\rm M} > 0$.}\label{}
\end{center}
\end{figure}

In Figure 10 we plot the particle count of the MS and LA as a function of Magellanic longitude.  The DB11 pure tidal model (solid line) exhibits ``plateaus" in contrast to the peaks of the GN96 model.  The impact of drag (dashed line) on the particle count of the MS is in fact \emph{opposite} to the case of GN96, because the drag actually \emph{creates} a peak in the particle count.  Considering the observation of a low density MS tip (Putman et al. 2003a), this effect of drag on the DB11 model is very much unwanted.  In contrast, drag has a desired effect on the particle counts of the LA.  The DB11 pure tidal model has a ratio of around two-to-one for particles in the MS as compared to the LA, but observations imply a more likely ratio of around four-to-one (Bruns et al. 2005).  By reducing the total particle count of the LA by more than half (Figure 10), the presence of drag is able to bring the total mass ratio between the MS and LA closer to the observed quantity.

\begin{figure}
\begin{center}
\includegraphics[width=8cm]{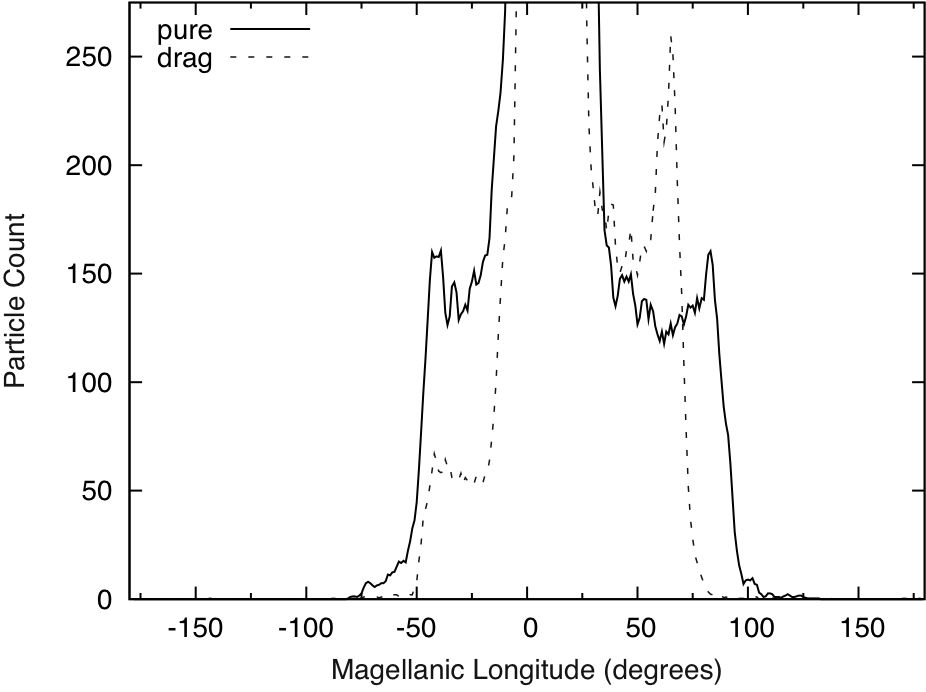}
\caption{Total number of particles in the pure tidal DB11 model (solid line) and the DB11 model with drag (dashed line) plotted as a function of Magellanic longitude.}\label{}
\end{center}
\end{figure}


\section{Understanding $\alpha$}

\subsection{Association with Hot Halo Density}

In the previous section, we found reasonably good ``tidal plus drag" models for the values of $\alpha =0.3$ (for GN96) and $\alpha =0.6$ (for DB11).  Even though $\alpha$ essentially scales the density of the hot halo, we have explained in section 2 that a given value for $\alpha$ does \emph{not} correspond to a unique value of the hot halo density.  The straightforward way to circumvent this problem is to utilize self-consistent hydrodynamical simulations, which would permit the hot halo density to be an independent parameter.  This approach is of course beyond the scope of the present study.  Instead, we provide a comparison with the results of Mastropietro et al. (2005, hereafter M05) in order to better understand the meaning of our adopted $\alpha$ values.

The ram pressure stripping model of M05 is a robust study of the hydrodynamical and gravitational interaction of the LMC with the Milky Way.  As the LMC plunges through the Milky Way halo, M05 find that ram pressure is able to strip away gas from the LMC into a long trailing stream which resembles the MS.  In order to create this stripping of the LMC disk, M05 adopt an NFW profile (Navarro et al. 1996) for the hot halo with a density of $8.5 \times 10^{-5}$ cm$^{-3}$ at 50 kpc.  In comparison, the present study adopts an isothermal profile for the hot halo with a density that is concealed in the physically ambiguous parameter $\alpha$.  Nevertheless, if we can determine the $\alpha$ values for which the LMC disk is altered by drag, we can associate these values with the densities of M05 as a rough approximation.

In order to isolate the effect of drag on the LMC disk, we perform an idealized simulation in which the LMC plunges face-on along the z-axis toward the Milky Way.  We remove the presence of the SMC and we enforce equation (\ref{dec}) to operate in the z-direction only.  We stop the simulation once the LMC arrives at the present-day distance of $\sim$50 kpc, and we inspect the state of the disk for different values of $\alpha$.  Because our treatment is not hydrodynamical, we do not observe an elegant trail of gas as in M05.  Instead, we find that increasing the value of $\alpha$ has two effects.  First, the positions of the disk particles are incrementally pushed further away from the LMC, and second, the disk succumbs to bending, particularly at its edges.

We find that the above two effects give a pronounced deviation from the morphology of the drag-free disk for values of $\alpha > 4.0$.  This result is largely independent of the range of LMC velocities explored in this study: both large velocities $\sim$360 km s$^{-1}$ (e.g. DB11) and small velocities $\sim$300 km s$^{-1}$ (e.g. GN96) give a similar threshold value of $\alpha \approx 4.0$.  We assume that ram pressure stripping cannot occur for smaller values of $\alpha$, and we therefore provide a rough association of $\alpha \approx 4.0$ with the M05 density of $8.5 \times 10^{-5}$ cm$^{-3}$ at 50 kpc.

The above estimate is a lower bound, for two reasons.  First, $\alpha \approx 4.0$ gives the \emph{threshold} value at which drag begins to alter the disk, but larger values are necessary to create robust stripping akin to M05.  And second, the NFW hot halo profile adopted by M05 falls off as $r^{-3}$, whereas our adopted isothermal halo falls off as $r^{-2}$.  Thus, even though we compare with the M05 density at $r=50$ kpc, the LMC in our simulation is subjected to higher densities at $r>50$ kpc, and the effect of drag is overestimated with respect to M05. 

The best ``tidal plus drag" models of the previous section utilize values of $\alpha$ which are approximately an order of magnitude less than $\alpha \approx 4.0$.  Accordingly, we suggest that these best models correspond to hot halo densities of $\sim 10^{-5}$ cm$^{-3}$ or less.

\subsection{Exclusivity of Tidal and Ram Pressure Origin}

When we simulate the SMC (rather than the LMC) plunging toward the Milky Way along the z-axis, we find that the threshold value for altering the SMC disk according to the aforementioned criteria is $\alpha \approx 3.0$.  Because this is less than the threshold value for the LMC, we conclude that the SMC should be stripped via ram pressure if the conditions for stripping the LMC are satisfied.  This makes sense because the SMC is less massive and is therefore less able to oppose the ram pressure exerted on its gas from the hot halo.  The M05 model is curious in this context, because it completely ignores the SMC, even though their model would likely support ram pressure stripping of the SMC disk.  Accordingly, we suggest that the MS formation scenario proposed by M05 should be revised to include contributions from the SMC gas.

The threshold value $\alpha \approx$ 3.0 for the SMC disk is significantly higher than the $\alpha$ values of the best ``tidal plus drag" models in section 4.  In fact, the respective values differ by an order of magnitude.  We therefore suggest that the hot halo densities in our best models are too small to support ram pressure stripping of the SMC disk.  This ``suggestion" can be made an ``assertion" if we adopt a full hydrodynamical treatment, and this will be the subject of future work.  Nevertheless, we need not abandon the present models in order to expand upon this idea.

For instance, consider two of the effects discussed in section 3: the sinking of the LA to lower Galactic radii (evident in the GN96 model), and the shortening of the MS (evident in the DB11 model).  Both of these effects increase in magnitude as $\alpha$ increases, and even though it is not immediately clear from section 3, both the GN96 model and the DB11 model suffer each of these effects.  At certain values of $\alpha$, the effect is so pronounced that the MS and LA are no longer identifiable.  Indeed, these tidal structures are ``destroyed" under the influence of drag: the MS progressively shortens until it becomes confined to the SMC disk, and the LA sinks deeper and deeper into the Galactic potential until much of it settles at the Galactic center.

For the DB11 model, we find that the MS are LA are destroyed under drag at a value of $\alpha \approx 2.0$, whereas the MS and LA in the GN96 model are destroyed at even smaller values of $\alpha$.  This suggests that the hot halo densities which disrupt the MS and LA are dangerously close to the densities of our ``best" models, possibly within a factor of only $\sim 3$.  This is not encouraging, as it implies that the MS and LA cannot survive in moderately dense $\sim 5 \times 10^{-5}$ cm$^{-3}$ hot halos.  This estimate is particularly disquieting in the context of our previous suggestion that our best models require a hot halo of density $\sim 10^{-5}$ cm$^{-3}$ in order to correct the LA kinematics, etc.  It would appear that ``fine-tuning" is necessary, which threatens the robustness of our models.

Though the MS and LA are destroyed for $\alpha \approx 2.0$, we stated previously that ram pressure stripping could occur for values as low as $\alpha \approx 3.0$ and $\alpha \approx 4.0$ for the SMC and LMC, respectively.  Thus, our results suggest that the structures which originate from tidal stripping are unable to survive at the hot halo densities required for ram pressure stripping.  If true, this means that tidal stripping and ram pressure stripping are mutually exclusive candidates for the formation of the MS and LA, because they require incompatible densities for the hot halo.  We can now assemble our knowledge of $\alpha$ and its effect on the MS and LA into Table 1, which organizes the various MS (and LA) formation scenarios as a function of $\alpha$ and the suggested hot halo density.  The suggested densities of Table 1 are taken at a distance of 50 kpc, and the estimates are based on our comparison with the results of M05.

\begin{table*}[t]
\begin{minipage}[c]{16cm}
\caption{Dependence of MS formation scenarios on $\alpha$ and $\rho$(50 kpc), the hot halo density at 50 kpc.}
\begin{tabular}{ccl}
\hline
$\alpha$ & Suggested $\rho$(50 kpc) (cm$^{-3}$) & Implied Model \\
\hline
0 & 0 & Pure tidal stripping model (e.g., GN96; DB11) \\
0.3-0.6 & $\sim 10^{-5}$ or less & Tidal plus drag model (e.g., this study) \\
2.0 & $\sim 5 \times 10^{-5}$ & Upper limit for survival of tidal features against drag \\
4.0 & $\sim 10^{-4}$ & Ram pressure stripping model (Mastropietro et al. 2005) \\
\hline
\end{tabular}
\end{minipage}
\end{table*}


\section{Discussion and Conclusions}

In this study we have analyzed the effect of hot halo drag on the global properties of the MS and LA, specifically in the context of the GN96 and DB11 tidal formation scenarios.  We have found that the drag creates a variety of both desired and undesired effects.  Comparing between the GN96 and DB11 cases, we furthermore find that the impact of drag isn't always consistent.  For instance, the peak in particle density at the MS tip is either removed by drag (GN96 case) or created by drag (DB11 case).  This suggests that our conclusions on the effect of drag are unfortunately model-dependent.

Nevertheless, our models agree on a few basic points.  For instance, the LA kinematics can be improved upon without significantly altering the morphology and kinematics of the MS.  In the GN96 case, the global properties of the MS remain largely unchanged, with only its column density being altered significantly.  In contrast, the drag forces the LA to sink to smaller Galactic radii, and the kinematics and on-sky position of the LA both change drastically.  In the DB11 model, drag causes the LA kinematics to agree much better with observations while the MS kinematics are negligibly adjusted.  We accordingly arrive at the conclusion that drag is able to influence the LA more strongly than the MS in our models.

Regardless of the specific model, we can assert that the hot halo drag will always impact the LA more strongly than the MS.  The reason is two-fold: the LA has a larger velocity than the MS when it is stripped from the SMC disk, and the LA probes denser regions of the hot halo.  The hot halo density $\rho$ will of course decrease with radius, and in the present study we have taken the specific case of an isothermal profile in equation (\ref{rho}).  Observations indicate that the LA resides at small Galactic radii $\sim$17 kpc (McClure-Griffiths et al. 2008), whereas the MS is traditionally assumed to lie at larger distances of $\sim$55 kpc (e.g., Putman et al. 2003a; Bruns et al. 2005).  Additionally, our results in Figures 4 and 9 suggest that the LA originates at smaller Galactic radii than the MS.  Accordingly, the LA is embedded in a higher density region of the hot halo and will suffer an increased ram pressure by an increase of $\rho$ in equation (\ref{rp}).

Because the LA leads the orbit of the SMC, it will naturally have larger galactocentric velocities than the MS during its evolution.  That is, in the process of elongating away from the SMC disk, the trailing features (i.e., the MS) must develop \emph{smaller} orbital velocities than the SMC, whereas leading features (i.e., the LA) must have \emph{larger} orbital velocities than the SMC.  The velocity relative to the hot halo can indeed be taken as the galactocentric (i.e. orbital) velocity of the MS and LA.  Equation (\ref{rp}) indicates that the ram pressure scales as the square of the galactocentric velocity $v^2$, and therefore the LA is subjected to greater drag than the MS as soon as it separates from the SMC disk.

In the GN96 model, the LA is dragged to very small Galactic radii ($\sim$20 kpc and less).  As stated in section 3, the new on-sky position of the dispersed LA does not correspond to any HI features of the Magellanic system, but we suggest that it may possibly correspond to other HI features of the Milky Way.  For instance, the distance, on-sky location, and velocity of the High Velocity Cloud Complex C are reasonably close to that of the dispersed LA in the GN96 drag model (Wakker 2001).  The suggestion that Complex C may have formed from in-falling HI gas is not new (Wakker et al. 1999), but its possible association with the Magellanic Clouds and specifically the LA has not been previously explored.   More extensive modeling should be able to indicate whether this formation scenario for Complex C is worthy of serious consideration.

The radial distance to the LA is constrained by only one data point, which was derived from an interaction of the LA with the Galactic disk (McClure-Griffiths et al. 2008).  The data point may have a large and furthermore unknown error bar, as the kinematic distance of $\sim$17 kpc is only as accurate as the adopted Milky Way rotation curve.  A distance of $\sim$17 kpc is much less than the distances to the LA as predicted by the pure tidal models.  We have found that hot halo drag is able to reduce though not resolve the discrepancy.  Future models should be motivated by the work of McClure-Griffiths et al. (2008) by incorporating the possibility of impacts between the LA and Milky Way disk, which would almost certainly have a significant impact on the LA morphology and kinematics (Bekki et al. 2008).

Though the origin of bifurcation within the MS has been the subject of speculation in the past (Putman et al. 2003a), we have presented a numerical model which convincingly reproduces the bifurcation from the interplay of gravitational and drag forces.  Moreover, we have proposed a formation mechanism which is sensitive to the magnitude of drag.  Without drag, the velocity dispersion of the MS is too large for the internal structure to condense into distinct filaments, and with too much drag, the MS itself is unable to survive.  The MS bifurcation may therefore be a sensitive probe of the magnitude of drag induced by the hot halo.  Even though the DB11 drag model reproduces the MS bifurcation, it also exhibits a number of obvious flaws, including the shortening of the MS and the undesired peak in particle density at its tip.  Accordingly, our DB11 drag model does not provide a complete understanding of the formation and evolution of the MS.  Nevertheless, the model is highly suggestive of the relevant physical mechanisms which have shaped the MS.  

The survival of the MS and LA against drag is a critical issue which informs the density estimates of Table 1.  We must point out, however, that these estimates (and more generally, the entire discussion of section 4) depends on the particular hot halo density profile that we have adopted.  If we instead choose a profile with a steeper slope at large radii, for instance the NFW profile, then the MS will be able to survive at greater values of the scale density $\rho_o$ (i.e., greater values of $\alpha$).  Accordingly, the conclusions of section 4 are model-dependent, relying in particular on our choice of an isothermal hot halo.

Even though the NFW profile may be more realistic, choosing anything other than an isothermal profile for the hot halo would violate the condition of hydrostatic equilibrium in the present study.  That is, our profile for the hot halo is predetermined by the profile of the dark matter halo utilized in the pure tidal models.  In order to properly study the evolution of the MS and LA in an NFW hot halo, we would need a tidal formation model which utilizes an NFW dark matter halo.  The effect of hot halo drag in such a model may very well differ from the results of the present study, and developing such a model will be the subject of future work.

The influence of the hot halo on the MS and LA is governed by a variety of gas-dynamical interactions, but the present study has simply assumed that the dominant effect is ram pressure drag.  Drag may indeed enforce the greatest changes in the global properties of the MS and LA, but the simple insertion of a drag term cannot fully address the complex interactions between the hot halo and the MS and LA.  For instance, neutral gas clouds can attain a multiphase structure via hydrodynamical interactions with the hot halo (Wolfire et al. 1995; Heitsch \& Putman 2009), and both the LA and the tip of the MS are observed to have such a multiphase structure (Bruns et al. 2005; Stanimirovic et al 2008).  Moreover, the column density distribution of the MS and LA are distinctly noncontinuous: the tip of the MS spreads out into fork and filaments (Nidever et al. 2010), and the HI clouds of the LA are disjointed and clumpy (Bruns et al. 2005).  This suggests a small-scale shaping mechanism which is ignored in the present study, such as thermal instabilities induced by hot halo interactions (Stanimirovic et al. 2008).

In addition, Westmeier \& Koribalski (2008) have found a large number of compact HI clouds that appear to be condensations of a largely ionized filament running alongside the MS.  The ionization of this purported filament, as well as the modest ionization of the MS itself, may plausibly be traced to collisions with hot halo gas (Sembach et al. 2003).  Such collisions may also explain the surprisingly large H-$\alpha$ emission observed within the MS (Putman et al. 2003b).  Bland-Hawthorn et al. (2007) have shown that the hot halo can slow down and disrupt the clouds of the MS as a prerequisite condition for inducing H-$\alpha$ emission at shock fronts.  It is therefore clear that the hot halo interacts with the MS and LA across a much more complex range of physics than considered in the present study.

Nevertheless, we have shown in the present study that the global properties of the MS and LA can change significantly under the influence of drag from the hot halo.  We have utilized the insertion of a simple drag term in our study, but a fully hydrodynamical treatment is needed in order to better understand the complete influence of the hot halo on the MS and LA.  We have shown that the tidal formation scenario can improve under the influence of drag, and we have also proposed a two-stage mechanism by which the MS bifurcation can forms.  Because these features are sensitive to the presence of drag, our results indicate that future modeling of the formation of the MS and LA will require a synthesis of tidal and drag forces.  In addition, our work suggests that the density of the hot halo may be constrained by considering the influence on the global properties of the MS and LA.

\section*{Acknowledgments}

JD is supported by a Sir Keith Murdoch Fellowship and a SIRF scholarship.  We thank the anonymous referee for criticism which led to the improvement of the manuscript.


\end{document}